\documentclass[twocolumn,aps,prl,nofootinbib]{revtex4-2}

\usepackage{amsmath,amssymb,bm}
\usepackage{graphicx}
\usepackage{microtype}
\usepackage{tabularx}
\usepackage[colorlinks=true,linkcolor=blue,citecolor=blue,urlcolor=blue]{hyperref}
\usepackage{orcidlink}
\newcommand{\Om}{\Omega_{D-2}}
\newcommand{\CV}{\Delta C_V}
\newcommand{\KD}{K_D}
\newcommand{\Kcal}{\mathcal K_D}
\newcommand{\dd}{\mathrm d}
\usepackage{xcolor}

\begin{document}

\title{Beyond the Thermodynamic-Volume Limit in Holographic Complexity}

\author{Di Wu\orcidlink{0000-0002-2509-6729}$^{1,2}$}
\email{wdcwnu@163.com, d277wu@uwaterloo.ca}

\author{Robert B. Mann\orcidlink{0000-0002-5859-2227}$^2$}
\email{rbmann@uwaterloo.ca}

\affiliation{$^1$School of Physics and Astronomy, China West Normal University,
Nanchong, Sichuan 637002, People's Republic of China  \\
$^2$Department of Physics and Astronomy, University of Waterloo, Waterloo, Ontario N2L 3G1, Canada}

\date{\today}

\begin{abstract}
In the large black hole regime, holographic complexity obeys a thermodynamic-volume scaling law, while the finite-size approach to this limit remains unresolved. In this Letter, we develop an analytic description of finite-size corrections to this scaling for complexity of formation in the complexity=volume (CV) proposal of odd-dimensional equal-spinning Myers-Perry-AdS black holes. We find that the leading finite-size correction retains state dependence that is absent from the limiting thermodynamic-volume law. This correction changes sign at a distinguished Critical Horizon Ratio (CHR), where the generic leading finite-size contribution vanishes and convergence to the thermodynamic-volume law is correspondingly accelerated. We further determine the dimension dependence of the CHR analytically, showing that it approaches a finite nonextremal value at large dimension with a calculable leading finite-dimensional correction. These results show that the finite-size approach can provide a new probe of holographic structure, motivating its exploration across different realizations of holographic complexity and its eventual characterization in the dual field theory.
\end{abstract}

\maketitle

Gauge/gravity duality encodes properties of strongly coupled quantum states in the geometry of a higher-dimensional spacetime \cite{ATMP2-231}. Entanglement entropy provides its best-understood realization, establishing a direct relation between quantum information and bulk geometry \cite{PRL96-181602,JHEP0707062}. Yet the bulk interior can continue to grow long after entanglement entropy has equilibrated. Holographic complexity was proposed as a probe of this continued bulk evolution \cite{FP64-24,FP64-49,PRD90-126007}. Among the standard holographic prescriptions, the complexity=volume (CV) proposal associates complexity directly with a maximal bulk volume, while the complexity=action (CA) proposal relates it to the action of a Wheeler-DeWitt patch \cite{PRD90-126007,PRL116-191301,PRD93-086006}.

A finite observable that isolates the state dependence is the complexity of formation, defined as the additional cost of preparing the thermofield-double state relative to two copies of the vacuum. For large static AdS black holes, it scales with the thermal entropy \cite{JHEP0117062}, whereas rotation separates entropy from thermodynamic volume. In extended black hole thermodynamics, the cosmological constant defines a pressure whose conjugate is the thermodynamic volume \cite{CQG26-195011,PRD84-024037,CQG34-063001}; for rotating AdS black holes, this volume is not fixed by the horizon area. For odd-dimensional equal-spinning Myers-Perry-AdS black holes, the large black hole CV complexity of formation instead exhibits a thermodynamic-volume scaling \cite{PRL126-101601,JHEP0521226},
\begin{equation}
 \CV\propto V^{\frac{D-2}{D-1}}.
 \label{eq:known-volume-law}
\end{equation}
Subsequent work extended the study of rotating holographic complexity across different gravitational and quantum settings \cite{JHEP1121037,PLB838-137691,JHEP0124055}. Generalized holographic complexity has also been studied for the equal-spinning Myers-Perry-AdS family, where the large black hole scaling was shown to persist beyond the standard prescriptions \cite{JHEP0924050}. The volume law establishes the large black hole limit, while leaving open how that limit is approached at finite size.

The approach to this limit can retain state dependence that is invisible in the limiting law itself. In this Letter, we determine this finite-size structure analytically for the CV complexity of formation of odd-dimensional equal-spinning Myers-Perry-AdS black holes. The leading correction changes sign at a distinguished Critical Horizon Ratio (CHR), where the generic finite-size contribution vanishes and convergence to the thermodynamic-volume law accelerates to order $\rho^{-2}$. The CHR furthermore approaches a simple analytic large-$D$ limit.

We use the equal-spinning Myers-Perry-AdS family as a controlled holographic laboratory in which the finite-size approach can be resolved analytically \cite{JHEP0521226,PRD59-064005,PRL93-171102,
PRD112-124024}. In odd spacetime dimension $D=2N+3$, let $r_+$ and $r_-$ denote the outer and inner horizon radii, and define $x\equiv r_-/r_+$ and $\rho\equiv r_+/\ell$, where $\ell$ is the AdS radius. We consider the large black hole limit $\rho\to\infty$ at fixed $0<x<1$; the endpoint limits $x\to0$ and $x\to1$ are separate and are not assumed to be uniform in this expansion. The metric functions relevant for the maximal-volume slice are  (see End Matter \cite{EM} for the full geometry)
\begin{align}
 F(r)\equiv g(r)^{-2}
 &=1+\frac{r^2}{\ell^2}
 -\frac{2m\Xi}{r^{D-3}}
 +\frac{B}{r^{D-1}}, \notag\\[-2pt]
 h(r)&=r\sqrt{1+\frac{B}{r^{D-1}}},
 \label{eq:metric-functions}
\end{align}
where $B\equiv2ma^2$, $\Xi\equiv1-a^2/\ell^2$, and $m$ and $a$ are the mass and rotation parameters. The horizons satisfy $F(r_+)=F(r_-)=0$; in the fixed-$x$ large horizon limit, these two conditions give
\begin{equation}
 B\simeq \frac{r_+^{D+1}}{\ell^2}b_D(x),
 \qquad
 b_D(x)=\frac{x^2(1-x^{D-1})}{1-x^2}.
\end{equation}
Defining $L\equiv B^{1/(D-1)}$, we then have
\begin{equation}
\lambda^{-1}\equiv  \frac{L}{r_+}\simeq
 b_D(x)^{1/(D-1)}\rho^{2/(D-1)}\gg1.
 \label{eq:scale}
\end{equation}
Hence the rotational contribution becomes $O(1)$ at an emergent radial scale $r\sim L$, parametrically outside the horizon. The quantity to be expanded is the vacuum-subtracted maximal volume,
\begin{align}
 \frac{G_NR}{2\Om}\CV
 &=\lim_{r_{\max}\to\infty}\Bigg[
 \int_{r_+}^{r_{\max}}\dd r\,h(r)r^{D-3}g(r) \notag\\[-2pt]
 &~~~-\int_0^{r_{\max}}\dd r\,
 \frac{r^{D-2}}{\sqrt{1+r^2/\ell^2}}\Bigg],
 \label{eq:CV-start}
\end{align}
where the asymptotically matched UV cutoff is understood. Rescaling the exterior region as $r=Lz$, we have $B/r^{D-1}=z^{-(D-1)}=O(1)$, while the remaining black hole terms in $F(r)$ are subleading at fixed $z$, so that $g(r)=\ell/r\,[1+o(1)]$. The leading volume difference between the black hole and the AdS background is therefore governed by the dimension-dependent integral
\begin{equation}
 I_D=\int_0^\infty \dd z\,z^{D-3}
 \left[\sqrt{1+z^{-(D-1)}}-1\right].
\end{equation}
The same emergent scale also controls the thermodynamic volume, whose leading behavior is
\begin{equation}
 V\sim \frac{\Om}{2(D-2)}L^{D-1},
\end{equation}
where $\Om$ is the area of the unit $(D-2)$-sphere. Eliminating $L$ between the two leading expressions yields
\begin{equation}
 \CV\sim \KD V^{\frac{D-2}{D-1}}.
\end{equation}
The coefficient is fixed analytically as
\begin{align}
\KD=\frac{\Om}{D-2}
 \frac{\Gamma\!\left(\frac{D-3}{2(D-1)}\right)
       \Gamma\!\left(\frac1{D-1}\right)}{\sqrt\pi}
 \left[\frac{2(D-2)}{\Om}\right]^{\frac{D-2}{D-1}}
 \label{eq:KD}
\end{align}
suppressing the common overall CV normalization $1/(G_NR)$. Although the emergent scale $L$ retains the horizon-ratio dependence, it drops out of the leading relation between $\CV$ and $V$.

\begin{figure}[t]
\includegraphics[width=\columnwidth]{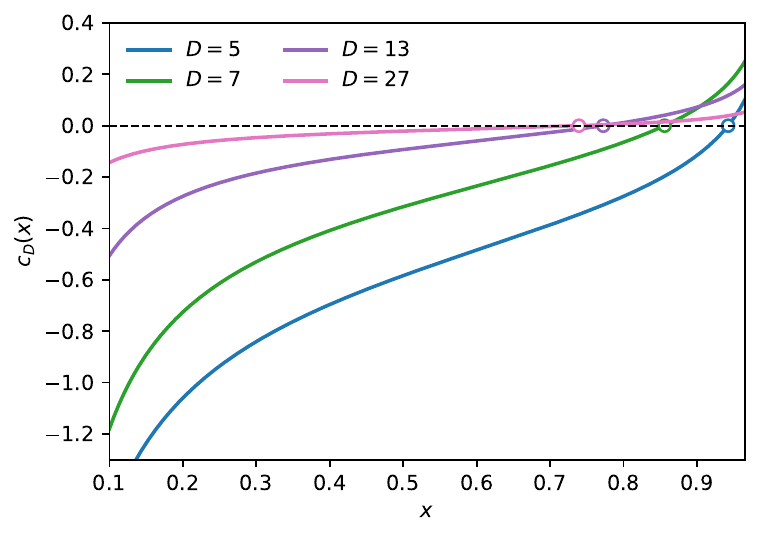}
\caption{State-dependent finite-size coefficient $c_D(x)$ for representative odd dimensions. Each curve crosses the dashed line $c_D=0$ at a dimension-dependent value $x_c$, marked by an open circle and defining the Critical Horizon Ratio (CHR).}
\label{fig:cD}
\end{figure}

The finite-size approach retains the state dependence erased from the leading law. Define $\Kcal(\rho,x)\equiv \CV/V^{(D-2)/(D-1)}$ and note from \eqref{eq:scale}  that $\lambda\ll1$. Since the exterior integral begins at $z_+=\lambda$ rather than at zero, its leading finite-size departure is controlled by the small-$z$ behavior of the outer integrand,
\begin{equation}
 z^{D-3}\left[\sqrt{1+z^{-(D-1)}}-1\right]
 =z^{\frac{D-5}{2}}-z^{D-3}+\cdots .
\end{equation}
The omitted outer contribution below the horizon therefore scales as
\begin{equation}
 \lambda^{\frac{D-3}{2}}\sim
 \rho^{-\frac{D-3}{D-1}}.
\end{equation}
The near-horizon region contributes at the same order, so the full coefficient requires matching an inner expansion in $y\equiv r/r_+$ to the outer expansion in $z=r/L$; see End Matter~\cite{EM} for details. This matched asymptotic expansion gives
\begin{align}
 \frac{\Kcal(\rho,x)}{\KD}
 &=1+c_D(x)\rho^{-\frac{D-3}{D-1}}
 +o\!\left(\rho^{-\frac{D-3}{D-1}}\right), \notag\\
 c_D(x)&=\frac{J_D(x)}{I_D}
 b_D(x)^{-\frac{D-2}{D-1}}.
 \label{eq:firstcorr}
\end{align}
Here
\begin{align}
 J_D(x)=\sqrt{b_D(x)}\Bigg[&\int_1^\infty\!\dd y
 \left(\frac{y^{D-2}}
 {\sqrt{y^{D+1}-a_D(x)y^2+b_D(x)}}\right. \notag\\[-2pt]
 &\left.-y^{\frac{D-5}{2}}\right)-\frac{2}{D-3}\Bigg],
 \label{eq:JD}
\end{align}
with $a_D(x)=(1-x^{D+1})/(1-x^2)$. The limiting law is therefore independent of $x$, while its leading finite-size approach is encoded in the analytic, state-dependent coefficient $c_D(x)$. As shown in Fig.~\ref{fig:cD}, $c_D(x)$ crosses zero at a distinguished value $x_c$,
\begin{equation}
 \boxed{c_D(x_c)=0},
 \label{eq:CHR}
\end{equation}
which we call the \emph{Critical Horizon Ratio} (CHR).  Here ``critical'' refers to the vanishing of the leading finite-size correction, rather than to a thermodynamic phase transition.  The curves in Fig.~\ref{fig:cD} further reveal that $c_D(x)$ changes sign smoothly from negative to positive across the CHR;
at $x=x_c$, the generic leading correction vanishes entirely.

\begin{figure*}[t]
\includegraphics[width=\textwidth]{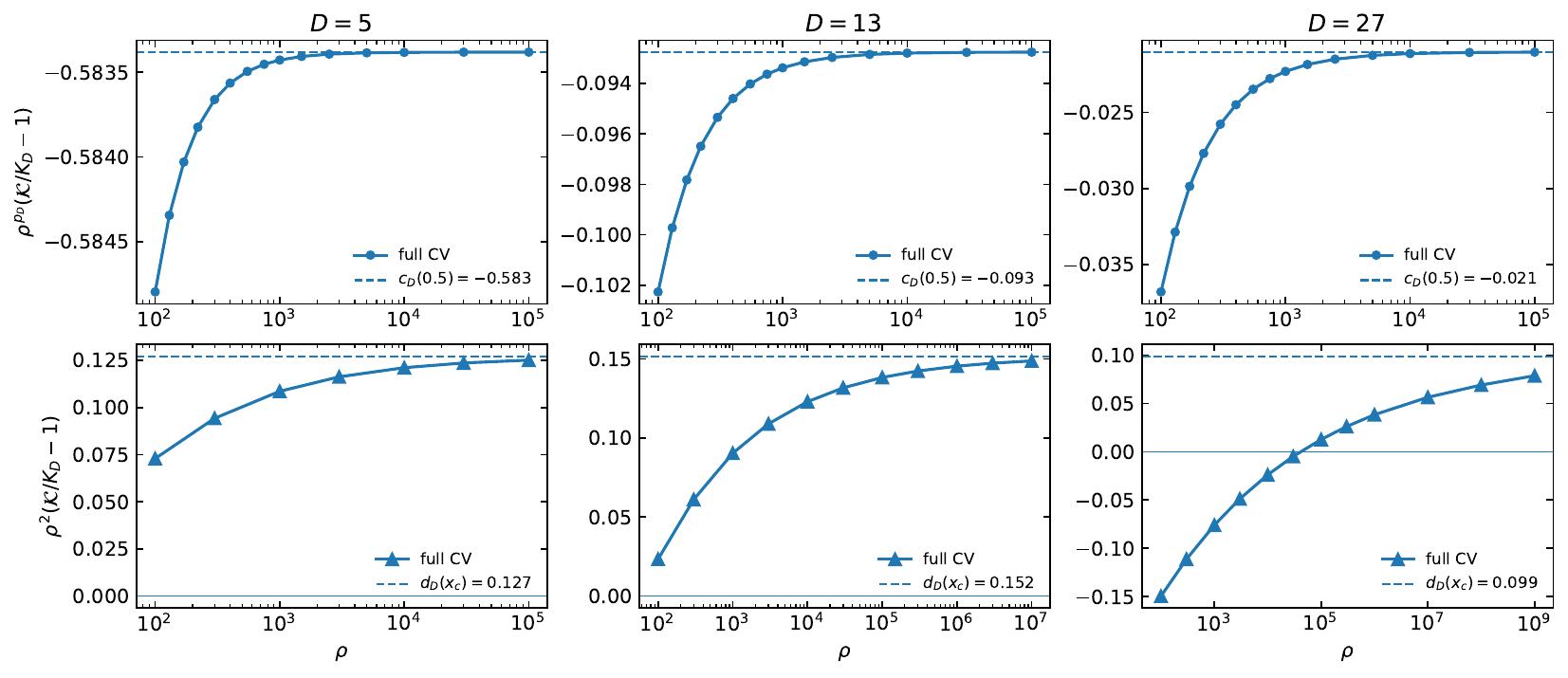}
\caption{Full-CV numerical tests of the finite-size expansion for representative dimensions $D=5$, $13$, and $27$.  \textbf{Top:} Away from the CHR ($x=0.5$), rescaling the finite-size correction by $\rho^{p_D}$, with $p_D=(D-3)/(D-1)$, yields convergence to the analytic coefficient $c_D(0.5)$ (dashed lines).  \textbf{Bottom:} At the CHR, rescaling by $\rho^2$ yields convergence to the predicted coefficient $d_D(x_c)$ (dashed lines).  The extended range for $D=27$ displays its slower approach to the asymptotic regime.}
\label{fig:scaling}
\end{figure*}

At the CHR, the generic $O(\lambda^{(D-3)/2})$ correction vanishes, requiring us to resolve the next order in the finite-size expansion.  The next term in the outer expansion produces an apparent correction of order $\lambda^{D-2}$.  At this order, the outer and interior-AdS contributions are
\begin{equation}
\delta C_V^{\rm out}=\frac{2\Omega_{D-2}\ell}{D-2}r_+^{D-2},~~~
\delta C_V^{\rm AdS}=-\frac{2\Omega_{D-2}\ell}{D-2}r_+^{D-2},
\end{equation}
so these terms cancel exactly and the contribution at order $\lambda^{D-2}$ vanishes~\cite{EM}.  Since
\begin{equation}
 \lambda\simeq b_D(x)^{-1/(D-1)}\rho^{-2/(D-1)},
\end{equation}
the first surviving algebraic correction at the CHR appears at order
\begin{equation}
 \lambda^{D-1}\simeq b_D(x_c)^{-1}\rho^{-2}.
\end{equation}
The CHR convergence law therefore takes the form
\begin{equation}
 \frac{\Kcal(\rho,x_c)}{\KD}
 =1+d_D(x_c)\rho^{-2}+\cdots .
 \label{eq:accelerated}
\end{equation}
Here $d_D(x_c)$ is fixed analytically by the next matched order~\cite{EM}, independently of the numerical evaluation below. Independent numerical evaluation of the full, unexpanded CV integrals confirms both asymptotic regimes: away from the CHR, the rescaled correction approaches the analytic coefficient $c_D(x)$, while at the CHR the $\rho^{-2}$ correction approaches its predicted coefficient, as shown in Fig.~\ref{fig:scaling}~\cite{EM}.

We next determine the large-$D$ behavior of the CHR analytically; this can be obtained by requiring $J_D$ in \eqref{eq:JD} to vanish in this limit.  Set $n=D-1$ and $b=x^2/(1-x^2)$.  At fixed $0<x<1$, $b_D(x)=b+O(x^n)$, while the exact identity $a_D(x)=1+b_D(x)$ gives $a_D(x)=1+b+O(x^n)$.
In the large-$n$ limit, the CHR integral is controlled by a boundary layer near $y=1$, with $y-1=O(n^{-1})$.  Introducing $y=e^{s/n}$ resolves this region while keeping $s=O(1)$.  The square-root denominator is then governed by
\begin{equation}
 Q_n(s)\equiv 1-(1+b)e^{-s}+b e^{-s}e^{-2s/n},
\end{equation}
whose large-$n$ expansion is
\begin{equation}
 Q_n(s)=1-e^{-s}-\frac{2bs e^{-s}}{n}
 +\frac{2bs^2 e^{-s}}{n^2}+O(n^{-3}).
\end{equation}
The leading boundary-layer structure is independent of $b$, so the horizon-ratio dependence first enters at subleading order.  Expanding the boundary-layer integral gives
\begin{equation}
 \mathcal J_n(b)=\frac{2}{n}
 +\frac{4+2\pi(b-1)}{n^2}
 +\frac{C_2(b)}{n^3}+\cdots,
\end{equation}
where $C_2(b)$ denotes the coefficient at third order in the boundary-layer expansion. The remaining term subtracted in the square brackets in Eq. \eqref{eq:JD} is
\begin{equation}
 \frac{2}{n-2}=\frac{2}{n}+\frac{4}{n^2}+\frac{8}{n^3}+\cdots.
\end{equation}
The $O(n^{-1})$ terms cancel and therefore do not constrain $b$.  At $O(n^{-2})$, the CHR implies
\begin{equation}
2\pi(b_\infty-1)=0,
\end{equation}
and hence
\begin{equation}
 b_\infty=1,\qquad x_c(\infty)=\frac{1}{\sqrt2}.
\end{equation}
The next order determines how the CHR approaches this limiting value at finite $D$.  We therefore write
\begin{equation}
 b_c= b_\infty+ \frac{\beta}{n}+O(n^{-2}),
\end{equation}
where $\beta$ denotes the leading finite-$n$ correction to $b_c$.  The sum of the $O(n^{-3})$ terms yields
\begin{equation}
 2\pi\beta+C_2(1)-8 =0.
\end{equation}
As evaluated explicitly in~\cite{EM}, $C_2(1)=8-16\pi\ln2$, and hence $\beta=8\ln2$. Using $x=\sqrt{b/(1+b)}$, this translates back to the horizon ratio as
\begin{equation}
 \boxed{x_c(D)=\frac1{\sqrt2}+\frac{\sqrt2\ln2}{D-1}
 +O\!\left((D-1)^{-2}\right)}.
 \label{eq:largeD}
\end{equation}
The CHR therefore remains a finite, nonextremal horizon ratio at large $D$, with both its limiting value and its leading finite-dimensional approach fixed analytically.

Taken together, these results show that the approach to the thermodynamic-volume limit contains structure that is invisible in the limiting law itself. While the leading relation between complexity of formation and thermodynamic volume is independent of the horizon ratio, the finite-size corrections retain this state dependence and organize the approach to the limit. The CHR exposes this structure most sharply: at this distinguished state, the generic leading correction vanishes and the asymptotic hierarchy itself is reorganized, accelerating convergence to the thermodynamic-volume law. The large-$D$ analysis further reveals that the CHR possesses an analytic structure of its own, approaching a finite nonextremal horizon ratio along a calculable trajectory. The finite-size expansion therefore does more than quantify deviations from the limiting law: it reveals an additional layer of structure in the approach of holographic complexity to its large black hole regime.

This additional structure raises a broader holographic question: which features of the finite-size approach are tied to a particular bulk realization, and which persist across different realizations of holographic complexity? Unequal-spin geometries, charged black holes, and other rotating AdS families provide natural laboratories in which this question can be tested by tracking how the finite-size hierarchy changes as the bulk realization is varied. The persistence of thermodynamic-volume scaling across different holographic settings then suggests a sharper question: does the finite-size approach itself exhibit a corresponding degree of universality? Our analysis in this Letter provides concrete quantities with which this possibility can be tested: the state-dependent finite-size correction, its distinguished zero at the CHR, and the associated reorganization of the asymptotic hierarchy. The resulting question is whether universality in holographic complexity extends beyond the limiting law to the structure of the approach itself.

A still deeper question is whether the finite-size structure identified on the bulk side admits an independent characterization in the dual strongly coupled theory. In particular, the CHR singles out a distinguished bulk state at which the leading finite-size correction vanishes and the asymptotic approach to the thermodynamic-volume law is reorganized. It is then natural to ask whether this same state can be identified directly within the boundary theory itself, without reference to the bulk construction that defines it here. Such an independent characterization would establish a direct connection between the finite-size organization of holographic complexity in the bulk and a distinguished structure in the dual field theory. The possibility of such a boundary counterpart therefore raises a broader question about the role of finite-size structure in the holographic dictionary of complexity.

\begin{acknowledgments}
This work is supported by the National Natural Science
Foundation of China (NSFC) under Grants No. 12675077 and No. 12375053, by the
China Scholarship Council (CSC), by the Sichuan Science and Technology Program
under Grant No. 2026NSFSC0021, and by the Natural Sciences and Engineering
Research Council of Canada.
\end{acknowledgments}

\twocolumngrid
\setcounter{equation}{0}
\renewcommand{\theequation}{A\arabic{equation}}
\renewcommand{\theHequation}{A.\arabic{equation}}
\setcounter{table}{0}
\renewcommand{\thetable}{A\arabic{table}}
\renewcommand{\theHtable}{A.\arabic{table}}

\onecolumngrid
\begin{center}
{\large\bfseries End Matter}
\end{center}
\twocolumngrid

\textit{Geometry and finite-size expansion.---}
For completeness, the odd-dimensional equal-spinning Myers-Perry-AdS geometry used in the Letter is
\begin{align}
 ds^2={}&-f(r)^2dt^2+g(r)^2dr^2+r^2\hat g_{ab}dx^a dx^b\notag\\[-2pt]
 &+h(r)^2[d\psi+\mathcal A-\Omega(r)dt]^2, \label{eq:E-metric}\\[-2pt]
 g(r)^2={}&\left(1+\frac{r^2}{\ell^2}-\frac{2m\Xi}{r^{D-3}}
 +\frac{2ma^2}{r^{D-1}}\right)^{-1},\notag\\[-2pt]
 h(r)^2={}&r^2\left(1+\frac{2ma^2}{r^{D-1}}\right),\qquad
 \Omega(r)=\frac{2ma}{r^{D-3}h(r)^2},\notag\\[-2pt]
 f(r)={}&\frac{r}{g(r)h(r)},\qquad
 \Xi=1-\frac{a^2}{\ell^2}. \label{eq:E-functions}
\end{align}
Here $\hat g_{ab}$ is the Fubini-Study metric on $\mathbb{CP}^{N}$, $D=2N+3$, normalized by $\mathrm{Ric}(\hat g)=2(N+1)\hat g$, and $d\mathcal A=2J$ with $J$ its K\"ahler form.  Defining $A\equiv2m\Xi$ and $B\equiv2ma^2$, the two horizons obey
\begin{equation}
 r_\pm^{D-1}+\frac{r_\pm^{D+1}}{\ell^2}-A r_\pm^2+B=0,
 \label{eq:E-horizon-poly}
\end{equation}
which fixes $A$ and $B$ exactly from $(r_+,r_-)$.  The thermodynamic volume is
\begin{align}
 V=\frac{\Omega_{D-2}r_+^{D-1}}{D-1}
 +\frac{\Omega_{D-2}B}{2(D-2)},\label{eq:E-volume}
\end{align}
Together with the vacuum-subtracted CV integral in Eq.~\eqref{eq:CV-start}, Eqs.~\eqref{eq:E-metric}-\eqref{eq:E-volume} provide the complete geometric
starting point for the analysis.

Setting $x\equiv r_-/r_+$ and $\rho\equiv r_+/\ell$, we introduce
\begin{equation}
 a_D(x)=\frac{1-x^{D+1}}{1-x^2},\quad
 \lambda=\frac{r_+}{L}\sim b_D^{-1/(D-1)}\rho^{-2/(D-1)}.
 \label{eq:E-lambda}
\end{equation}
At fixed $0<x<1$, the horizon equations at large $\rho$ give $A\simeq r_+^{D-1}a_D/\ell^2$ and $B\simeq r_+^{D+1}b_D/\ell^2$. Setting $y=r/r_+$, the leading inner integrand is proportional to
\begin{equation}
 \sqrt{b_D}\,\frac{y^{D-2}}{\sqrt{y^{D+1}-a_Dy^2+b_D}},
\end{equation}
whose large-$y$ overlap with the outer region is $\sqrt{b_D}\,y^{(D-5)/2}$. Subtracting this overlap and restoring its analytically integrated finite part, $-2\sqrt{b_D}/(D-3)$, yields precisely $J_D(x)$ in Eq.~\eqref{eq:JD}. Together with the universal outer integral $I_D$, this reproduces Eq.~\eqref{eq:firstcorr}. The CHR condition $c_D(x_c)=0$ is therefore equivalently $J_D(x_c)=0$.

The missing outer interval contains
\begin{equation}
 \int_0^\lambda dz\,z^{D-3}[\sqrt{1+z^{-(D-1)}}-1]
 =\frac{2\lambda^{(D-3)/2}}{D-3}-\frac{\lambda^{D-2}}{D-2}+\cdots .
\end{equation}
The second term gives $+\ell r_+^{D-2}/(D-2)$, whereas the interior AdS subtraction gives
\begin{equation}
 -\int_0^{r_+}\frac{r^{D-2}dr}{\sqrt{1+r^2/\ell^2}}
 =-\frac{\ell r_+^{D-2}}{D-2}+O(\ell^3r_+^{D-4}),
\end{equation}
so $[\Delta C_V]_{\lambda^{D-2}}=0$ for every fixed $0<x<1$.  At the CHR the earlier $\lambda^{(D-3)/2}$ term also vanishes.  The first surviving term is $\lambda^{D-1}\sim\rho^{-2}$.  Its matched finite part is
\begin{align}
 H_D&=\operatorname{FP}\!\int_0^\infty z^{-2}\sqrt{1+z^{-(D-1)}}dz \notag\\[-2pt]
 &=\frac{\Gamma(\frac1{D-1})\Gamma(-\frac12-\frac1{D-1})}
 {(D-1)\Gamma(-1/2)}.
\end{align}
Here $\operatorname{FP}$ denotes the Hadamard finite part obtained after subtracting the small-$z$ divergent terms.  This yields
\begin{equation}
 \boxed{d_D(x)=\frac{a_D(x)}{2b_D(x)}\frac{H_D}{I_D}
 -\frac{2(D-2)^2}{(D-1)^2b_D(x)}}.
 \label{eq:E-dD}
\end{equation}

\textit{Large-$D$ expansion and validity.---}
The step omitted from the Letter is the evaluation of $C_2(1)$. With $n=D-1$, $b=x^2/(1-x^2)$, and $y=e^{s/n}$,
\begin{align}
 C_2(1)=\int_0^\infty ds\,s^2e^{s/2}\Bigg[&\frac12\left(\frac1{\sqrt{1-e^{-s}}}-1\right)
 -\frac{2e^{-s}}{(1-e^{-s})^{3/2}}\notag\\[-2pt]
 &+\frac{3e^{-2s}}{2(1-e^{-s})^{5/2}}\Bigg].
\end{align}
A common endpoint regulator and $u=e^{-s}$ reduce the combination to second derivatives of Beta functions.  Digamma and trigamma recurrences cancel the regulator poles, Euler constants, and $\pi^2$ pieces, leaving
\begin{equation}
 \boxed{C_2(1)=8-16\pi\ln2},
\end{equation}
which gives $\beta=8\ln2$ in the main text matching condition.

The expansion is pointwise at fixed $0<x<1$ and is not uniform at either endpoint.  As $x\to0$, $b_D\sim x^2$ and scale separation requires $xr_+/\ell\gg1$; the static limit therefore cannot be obtained by direct substitution.  The near-extremal limit $x\to1$ is likewise separate.  A common Fefferman-Graham regulator gives
\begin{equation}
 r_{\max}^{\rm BH}-r_{\max}^{\rm AdS}
 =\frac{m\Xi}{(2N+2)\ell^{4N}}\delta^{2N+1}\to0,
\end{equation}
so no finite cutoff-matching term is required.

\textit{Numerical checks.---}
The unexpanded CV integral was evaluated directly, independently of the asymptotic coefficients. For each $(D,x,\rho)$ the exact horizon equations determine $A$ and $B$; the horizon square-root singularity is removed with $r=r_++t^2$, the integration is segmented, and the asymptotic tail is added analytically. Table~\ref{tab:E-leading} records all sixty leading-law checks at $\rho=300$.

\begin{table}[t]
\caption{Relative deviation $100(K_D^{\rm num}/K_D-1)$ (percent) at $\rho=300$; sixty independent full-CV checks.}
\resizebox{0.35\textwidth}{!}{
\begin{tabular}{crrrrr}
\hline\hline
$D$&$x=0.1$&$x=0.3$&$x=0.5$&$x=0.7$&$x=0.9$\\\hline
5&$-8.860$&$-4.857$&$-3.370$&$-2.230$&$-0.665$\\
7&$-2.714$&$-1.190$&$-0.706$&$-0.347$&$+0.151$\\
9&$-1.297$&$-0.498$&$-0.270$&$-0.106$&$+0.127$\\
11&$-0.777$&$-0.269$&$-0.137$&$-0.045$&$+0.088$\\
13&$-0.534$&$-0.168$&$-0.082$&$-0.024$&$+0.063$\\
15&$-0.401$&$-0.116$&$-0.055$&$-0.014$&$+0.046$\\
17&$-0.323$&$-0.087$&$-0.040$&$-0.010$&$+0.034$\\
19&$-0.271$&$-0.067$&$-0.030$&$-0.007$&$+0.027$\\
21&$-0.237$&$-0.056$&$-0.024$&$-0.006$&$+0.021$\\
23&$-0.211$&$-0.045$&$-0.019$&$-0.004$&$+0.018$\\
25&$-0.192$&$-0.039$&$-0.016$&$-0.003$&$+0.015$\\
27&$-0.186$&$-0.041$&$-0.020$&$-0.009$&$+0.006$\\
\hline\hline
\end{tabular}}
\label{tab:E-leading}
\end{table}

\begin{table}[h]
\caption{Analytic CHR data: $c_D(x_c)=0$ and Eq.~\eqref{eq:E-dD}.}
\resizebox{0.45\textwidth}{!}{
\begin{tabular}{c|rr|r|rr}
\hline\hline
$D$&$x_c$&$d_D$&$D$&$x_c$&$d_D$\\\hline
5&0.942235202&0.127090254&17&0.756165253&0.132983265\\
7&0.855735063&0.165157723&19&0.751195825&0.124667807\\
9&0.810998736&0.167821644&21&0.747297540&0.117136290\\
11&0.786767652&0.160985640&23&0.744122328&0.110342951\\
13&0.772276772&0.151642334&25&0.741461668&0.104217042\\
15&0.762816095&0.142052400&27&0.739184478&0.098683796\\
\hline\hline
\end{tabular}}
\label{tab:E-CHR}
\end{table}

For the upper panels of Fig.~\ref{fig:scaling}, the parameter-free targets are
\begin{align}
 c_5(0.5)&=-0.583381565, & c_{13}(0.5)&=-0.092753391,\notag\\[-2pt]
 c_{27}(0.5)&=-0.021031808.&&
\end{align}
The lower-panel targets are the corresponding entries in Table~\ref{tab:E-CHR}. The extended $D=27$ range is required because subleading terms delay the onset of the positive asymptotic $\rho^{-2}$ sector.

\end{document}